\documentclass[manuscript]{acmart}
\AtBeginDocument{%
  \providecommand\BibTeX{{%
    \normalfont B\kern-0.5em{\scshape i\kern-0.25em b}\kern-0.8em\TeX}}}

\setcopyright{acmcopyright}
\copyrightyear{2023}
\acmYear{2023}
\acmDOI{XXXXXXX.XXXXXXX}

\acmConference[QP 2023]{The First International Workshop on the Art, Science, and Engineering of Quantum Programming (QP)}{March 13--14,
  2023}{Tokyo, Japan}
\acmPrice{}
\acmISBN{}

\begin{document}

\title{Model-Driven Quantum Federated Learning (QFL)}


\author{Armin Moin}
\affiliation{%
  %\department{School of CIT}
  \institution{Technical University of Munich \& University of Antwerp, armin.moin@tum.de}
 \city{Munich \& Antwerp}
  \country{Germany \& Belgium}
  }
\email{armin.moin@tum.de}

\author{Atta Badii}
\affiliation{%
  %\department{Department of \\ Computer Science}
  \institution{University of Reading, atta.badii@reading.ac.uk}
  \city{Reading}
  \country{United Kingdom}
}
\email{atta.badii@reading.ac.uk}

\author{Moharram Challenger}
\affiliation{%
  %\department{Department of Computer Science}
  \institution{University of Antwerp \& Flanders Make, moharram.challenger@uantwerpen.be}
  \city{Antwerp}
  \country{Belgium}
  }
\email{moharram.challenger@uantwerpen.be}

\renewcommand{\shortauthors}{Moin et al.}

\begin{abstract}
  Recently, several studies have proposed frameworks for Quantum Federated Learning (QFL). For instance, the Google TensorFlow Quantum (TFQ) and TensorFlow Federated (TFF) libraries have been deployed for realizing QFL. However, developers, in the main, are not as yet familiar with Quantum Computing (QC) libraries and frameworks. A Domain-Specific Modeling Language (DSML) that provides an abstraction layer over the underlying QC and Federated Learning (FL) libraries would be beneficial. This could enable practitioners to carry out software development and data science tasks efficiently while deploying the state of the art in Quantum Machine Learning (QML). In this position paper, we propose extending existing domain-specific Model-Driven Engineering (MDE) tools for Machine Learning (ML) enabled systems, such as MontiAnna, ML-Quadrat, and GreyCat, to support QFL.
\end{abstract}


\keywords{model-driven engineering, quantum computing, federated machine learning}



\maketitle

\section{Introduction and related work}
\label{sec:intro}
Federated Learning (FL) enables a scalable, privacy-preserving-by-design approach to Machine Learning (ML) since raw data are not exchanged between distributed nodes running the learning algorithm for training the ML model. Hence, in each round, a server that stores the \textit{current} version of the ML model parameters shares these parameters with the distributed nodes, called \textit{clients}, that store the raw data required for ML model training. Each node uses its local data to run a learning algorithm, such as Stochastic Gradient Descent (SGD), and update its local model parameters accordingly. Further, each client sends its \textit{updated} ML model parameters to the server. Finally, the server applies federated averaging \cite{McMahan+2017} to calculate the new ML model parameters and shares these parameters with all of the clients \cite{ChehimiSaad2022}.

However, privacy preservation is not the only benefit of this decentralized ML approach. The fact that ML model parameters are exchanged rather than raw data optimizes the network throughput and would contribute to energy efficiency gains and, thus, carbon emissions footprint. This is particularly the case where the data objects consist of multi-variate structures, such as complex, high-resolution color image data, the transmission, and processing of which is typically associated with higher energy cost and carbon emissions \cite{RavelBadii2019}. The capability to deliver decentralized ML becomes even more crucial when we deal with quantum data with an inherently fragile nature, which makes them difficult to transfer. Note that some advanced Quantum Computing (QC) technologies, which are adequate for Quantum Machine Learning (QML), require special conditions, such as extremely low temperatures and vibration-free environments, to store qubits and effectively maintain their quantum states \cite{NielsenChuang2010}. This makes any exchange of quantum data between the distributed nodes running decentralized ML highly difficult and inefficient since quantum data decay gradually as they interact with the environment \cite{NielsenChuang2010, ChehimiSaad2022}. The advantage of Quantum Federated Learning (QFL) is that we can process inherently quantum data on the distributed, federated nodes training the ML model (i.e., clients) but use only classical (i.e., non-quantum) data for the ML model parameters, which will be exchanged.

Recently, Chehimi and Saad \cite{ChehimiSaad2022} proposed a framework for QFL. They deployed the Google TensorFlow Quantum (TFQ) and TensorFlow Federated (TFF) libraries. Additionally, Yun et al. \cite{Yun+2022} proposed SlimQFL to support QFL using Slimmable Neural Networks (SNNs), thus coping with environmental dynamics, such as time-varying communication channel conditions and energy limitations. However, the majority of practitioners in Software Engineering (SE) and Data Science (DS) are yet to become familiar with the Quantum Programming (QP) paradigm and the technologies mentioned above. Therefore, as highlighted by prior work in the literature, for example, Ali and Yue \cite{AliYue2020}, Delgado and Gonzalez \cite{DelgadoGonzalez2020}, Gemeinhardt et al. \cite{Gemeinhardt+2021}, and Moin et al. \cite{Moin+2022a}, modeling languages and the Model-Driven Engineering (MDE) paradigm could be a natural fit and beneficial for QP. However, none of them have pointed out the need for supporting model-driven QFL.

\section{Proposed approach}
\label{sec:proposed-approach}
We envision a future in the decade ahead in which hybrid quantum-classical applications will deploy QFL technologies to address the increasingly challenging enablers for AI solutions efficiently. However, as mentioned above, few practitioners (i.e., developers and data scientists) are currently familiar with QP, let alone QFL. Hence, to facilitate the deployment of QFL technologies in ML-enabled software-intensive systems, we propose extending existing DSMLs and MDE tools as enabling technologies to support the development of such systems. This requires enhancement of state-of-the-art tools, such as MontiAnna \cite{Kusmenko+2019}, ML-Quadrat \cite{Moin+2022b}, and GreyCat \cite{HARTMANN+2019a, Hartmann+2019b} to address QFL. 

Realizing the proposed approach will help increase the productivity of practitioners, shorten the time-to-market for new software-based products and services, and increase the quality of software systems. Domain-specific modeling has already been applied to other domains and has resulted in a productivity leap of 500-1000\% \cite{KellyTolvanen2008}. Moreover, the possibility of achieving full automation for program synthesis (i.e., generation of the source code and other artifacts, such as ML models) makes turnarounds for developing software-based products and services shorter and increases software quality, for example, by reducing the number of software defects.

\begin{figure}[ht]
	\centering
	\includegraphics[width=0.3\textwidth]{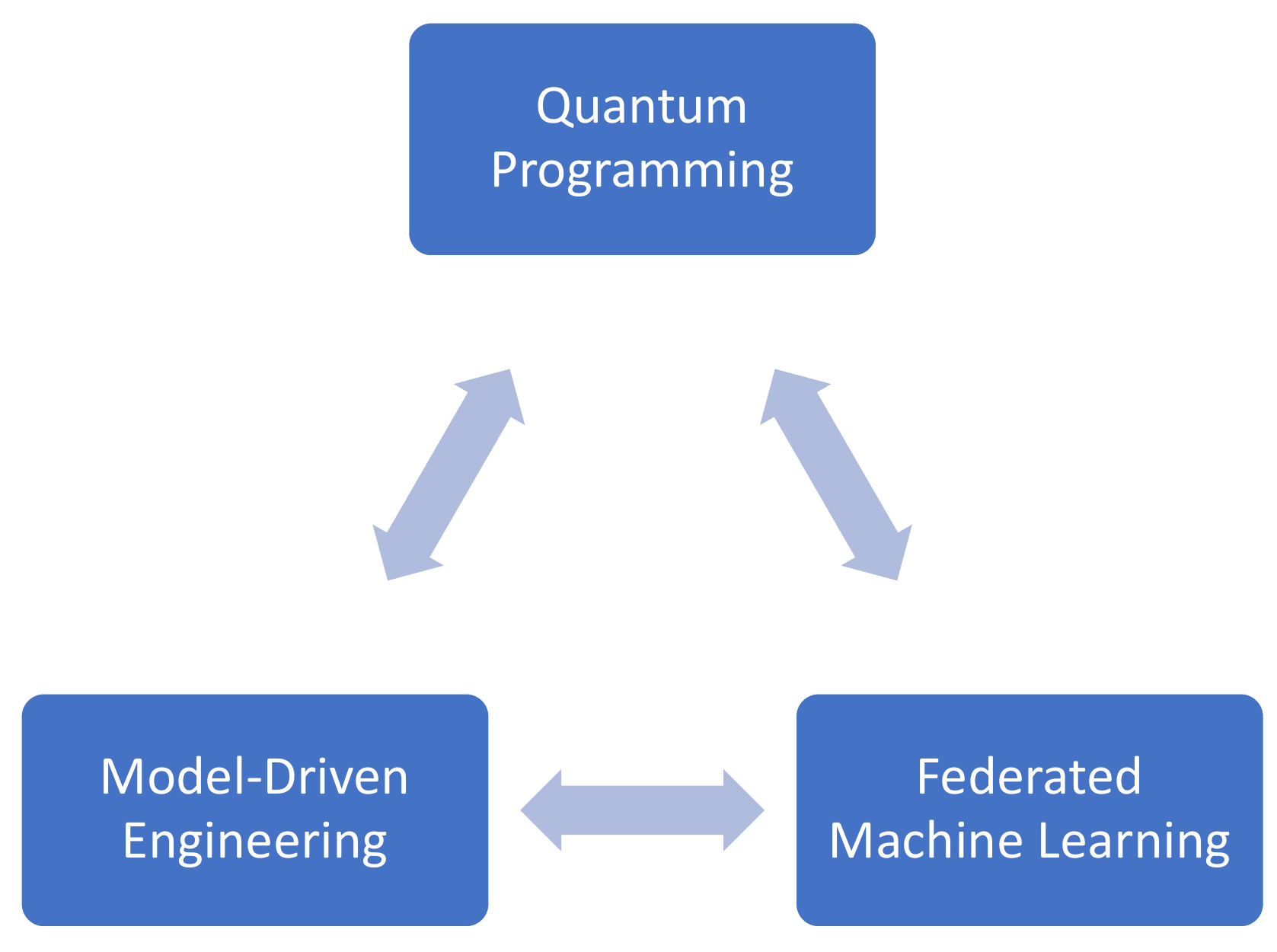}
	\caption{The main pillars of the proposed approach}
	\label{fig:topics}
\end{figure}

The DSML and modeling tool for delivering the above enabling technologies should support the user (i.e., the practitioner) to specify the number of distributed nodes (i.e., the clients and the server) participating in FL at the design time. Moreover, for each node, it should be specified whether this node - on the physical layer - possesses a quantum processor or a classical one. Further, the specific quantum hardware technology (e.g., superconducting NISQ vs. trapped-ions) and model of computation (e.g., quantum circuits vs. quantum annealing) may be provided. This should help optimize the generated code for the particular quantum and classical processors on which the QFL application will be deployed.

Additionally, it should be possible to choose the preferred learning algorithm for each client participating in FL, the desired federated averaging algorithm (e.g., appropriate aggregation of model gradients), and the ML model architecture. In each case, the system should be able to revert to a best-suited \textit{default} choice to be selected automatically should the user (i.e., practitioner) not provide the respective information explicitly. For instance, the default choices for the learning algorithm, federated averaging algorithm, and ML model architecture could be SGD, the federated averaging algorithm of \cite{McMahan+2017}, and Quantum Convolutional Neural Networks (QCNN) \cite{ChehimiSaad2022}, respectively.

In particular, we plan to integrate the open-source TFQ and TFF libraries based on Cirq \cite{Cirq}, thus building on prior work by Chehimi and Saad \cite{ChehimiSaad2022}. Furthermore, we intend to extend the open-source ML-Quadrat project \cite{Moin+2022b}, built based on the Eclipse Modeling Framework (EMF) and the Xtext framework. Currently, ML-Quadrat supports neither QP nor FL. Thus, an approach integrating both of the requisite capabilities (namely, QP and FL, thus QFL) will be required. Finally, the DSML and tool can be enhanced with more sophisticated QFL solutions, such as the SNN-based approach proposed by Yun et al. \cite{Yun+2022}. The DSML and tool should be modular and extensible to support the above requirements. Figure \ref{fig:topics} illustrates the main pillars of the proposed approach.

\section{Conclusion and future work}
\label{sec:conclusion}
In this paper, we have proposed our approach to delivering model-driven QFL. We have argued that existing DSMLs and modeling tools should be extended to handle advanced ML needs, specifically regarding QFL. In the future, we plan to realize the proposed solution and conduct experiments to validate the prototype. To this aim, we will build on existing QFL and MDE solutions.

%\begin{acks}
%To Robert, for the bagels and explaining CMYK and color spaces.
%\end{acks}

\bibliographystyle{ACM-Reference-Format}
\bibliography{refs}

%\appendix

\end{document}